\documentclass[prl,floatfix,showpacs,aps,reprint,letterpaper,nobalancelastpage]{revtex4-1}
\usepackage{graphicx} 
\usepackage{color} 
\usepackage{transparent}
\usepackage{amsmath}
\usepackage{hyperref} 
\usepackage{amssymb}

\begin{document}

\title{Machine learning for discriminating quantum measurement trajectories and improving readout}

\author{Easwar Magesan}
\affiliation{IBM T.J. Watson Research Center, Yorktown Heights, NY 10598, USA}
\author{Jay M. Gambetta}
\affiliation{IBM T.J. Watson Research Center, Yorktown Heights, NY 10598, USA}
\author{A.D. C\'orcoles}
\affiliation{IBM T.J. Watson Research Center, Yorktown Heights, NY 10598, USA}
\author{Jerry M. Chow}
\affiliation{IBM T.J. Watson Research Center, Yorktown Heights, NY 10598, USA}
\begin{abstract}
High-fidelity measurements are important for the implementation of quantum information protocols. Current methods for classifying measurement trajectories in superconducting qubit systems produce fidelities systematically lower than those predicted by experimental parameters. Here, we place current classification methods within the framework of machine learning (ML) algorithms and improve on them by investigating more sophisticated ML approaches. We find that non-linear algorithms and clustering methods produce significantly higher assignment fidelities that help close the gap to the fidelity achievable under ideal noise conditions. Clustering methods group trajectories into natural subsets within the data, which allows for the diagnosis of systematic errors. We find large clusters in the data associated with $\mathrm{T}_1$ processes and show these are the main source of discrepancy between our experimental and achievable fidelities. These error diagnosis techniques help provide a path forward to improve qubit measurements.
\end{abstract}
\maketitle

The ability to perform accurate measurements is important for maximizing the information one can extract from a physical system. This is especially true in experimental quantum information processing since quantum systems are highly susceptible to noise effects and error rates of quantum operations and measurements must be small for fault-tolerant quantum computation to be a reality~\cite{Sho96}. Our goal is to provide methods for diagnosing measurement errors and increasing fidelities by using classification and clustering algorithms in machine learning (ML). While we apply our methods in a superconducting qubit measurement system, we anticipate that the generality of these techniques can be useful in a broader class of systems.

Superconducting quantum bits (qubits) have become a promising candidate for building a fault-tolerant quantum computer due to their long coherence times~\cite{Paik2011,chang_improved_2013,Barends2013}, high-fidelity multi-qubit gate operations~\cite{Barends_superconducting_2014}, and the abilitiy to perform single-shot measurements~\cite{Mallet09,Bergeal10,Johnson12,Riste12}, in a circuit QED architecture~\cite{BHW04}. Remarkable progress has been made in reducing error-rates of these operations however considerable work is still needed to implement fault-tolerant quantum computation in large networks of qubits. In circuit quantum electrodynamics (cQED) a superconducting anharmonic oscillator, such as a transmon~\cite{Koch07}, is coupled to a resonator, producing a state-dependent shift of the resonator frequency. This allows for qubit measurements by driving the resonator and recording the output trajectory~\cite{gambetta_trajectories_2008} in phase (I-Q) space. In practice there are significant sources of random noise and systematic effects, such as $\mathrm{T}_1$ processes where the system spontaneously decays to its ground state, that can make single-shot trajectories appear complex and difficult to distinguish.

Our experimental system is a single qubit (Q4) in a planar lattice of four superconducting qubits~\cite{Corcoles2014}. We show current methods for assigning outcomes to measurement trajectories in this system produce assignment fidelities (defined below in Eq.~(\ref{eq:assfid})) that are much lower than the predicted achievable value derived under ideal noise conditions . We utilize various ML algorithms to obtain deeper insight into the behavior of the trajectories and bring fidelities up closer to our expected values. We find a total increase in assignment fidelity from 0.9586 using current methods to 0.9821 ($\sim 2.4\%$ increase) using non-linear ML classifiers. The strong performance of non-linear classifiers indicates systematic effects, such as heating and $\mathrm{T}_1$ events, could be a significant source of error in our measurements.

To verify this, we use ML clustering methods to group the data into naturally occurring subsets. We find a large cluster consisting of $\mathrm{T}_1$ events whose size is consistent with the experimentally measured $\mathrm{T}_1$ time. Replacing this cluster with random non-$\mathrm{T}_1$ events gives assignment fidelities approaching 0.995 which is much closer to the achievable value of 0.9999. Going to higher orders we find a much small cluster corresponding to heating of the ground state into the excited state. Before moving on, let us make a few points about using ML methods for trajectory discrimination and improving measurements. First, the methods we present here can be useful in a much broader context. Any measurement scheme that produces patterns in a geometric space can potentially benefit from more advanced ML methods. Investigating the applicability to different systems will depend on the details of each situation. Second, these methods are applicable even if we are trying to improve higher fidelity measurements than that of this paper. The key is that these methods can be tailored according to the types of noise present. Third, ML methods have also been applied to other problems in quantum information such as phase estimation~\cite{Hentschel2010} and asymptotic state estimation~\cite{Guta2010}.

The standard metric for characterizing how well a single-qubit measurement assigns outcomes is the assignment fidelity
\begin{equation}\label{eq:assfid}
\mathcal{F}_a = 1-\left(\mathbb{P}\left[0|1\right] + \mathbb{P}\left[1|0\right]\right)/2.
\end{equation}
Here $\mathbb{P}[0|1]$ ($\mathbb{P}[1|0]$) is the probability of obtaining outcome ``0" (``1") given the system was prepared in $|1\rangle$ ($|0\rangle$). Hence $\mathcal{F}_a \in [0,1]$ and ideally $\mathcal{F}_a \sim 1$. Our measurement framework is the dispersive limit of cQED, where observing the resonator output voltage provides a quantum non-demolition qubit measurement. For outcome ``0" (1) the voltage leaving the cavity represents a coherent state I-Q trajectory, and single-shot trajectories are obtained by amplifying the cavity signal. The main parameters of our system are given in~\cite{Supp} and complete experimental information is given in Ref.~\cite{Corcoles2014}.

\begin{figure}[h!]
\centering
\includegraphics[trim = 0.5in 2.0in 0.1in 2.0in,width=0.43\textwidth]{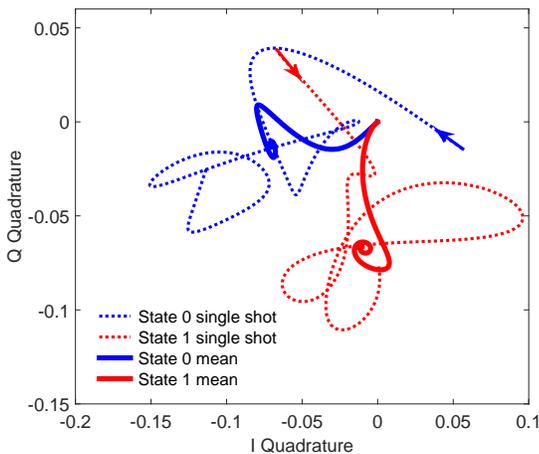}
\caption{\label{fig:Mean_trajs} Mean trajectories and single-shots for $|0\rangle$ (blue) and $|1\rangle$ (red) preparations (color online). The $|0\rangle$ ($|1\rangle$) single-shot trajectory (blue (red)-dotted) has arrow pointing up (down) and to the left (right). The mean trajectories of $|0\rangle$ (blue-solid) and $|1\rangle$) (red-solid) have steady-states of $\sim$ (-0.07,-0.02) and (-0.01,-0.07). }
\end{figure}

Our data consists of 51200 single-shot trajectories (shots), half initially prepared in $|0\rangle$ and the other half in $|1\rangle$ (denote these classes by $C_0$ and $C_1$). The trajectories are time-ordered and the first half is used as a training set to predict the second half for classification. The mean trajectories for each class, denoted $\mu_0(t)$ and $\mu_1(t)$, as well as examples of single-shot trajectories, are given in Fig.~\ref{fig:Mean_trajs}. We see that each shot can be complicated but there are enough shots to ensure smooth, well-separated means. The total measurement time $T$ is $2.6 \mu$s and $[0,T]$ is discretized into 163 time-points so trajectories are represented by vectors $x \in \mathbb{R}^{326}$ (let $M=326$) where the first (last) 163 entries correspond to the real (imaginary) parts of the trajectory.

The current method of classifying trajectories~\cite{ryan_inprep2013} is to integrate each trajectory with a filter (kernel,weight) function $w(j)$. Formally, if $x \in \mathbb{R}^M$ is a single trajectory, under the assumption that the covariance matrices $\Sigma_0$ and $\Sigma_1$ of each class are equal, Gaussian, and diagonal, the optimal $w(j)$ is equal to the ratio of the difference between the mean trajectories $\mu_0$ and $\mu_1$ and the variance $v$. If $\Sigma_d$ is the diagonal covariance matrix then
\begin{align}\label{eq:LDAdiag}
f(x)&=  x^T\left[\Sigma_d^{-1}\left(\mu_0-\mu_1\right)\right].
\end{align}
The assignment fidelity using this method is $\mathcal{F}_a =0.9586$.

 The ``achievable" assignment fidelity for our experiment, $\mathcal{F}_\mathrm{ach}$, is the fidelity we would obtain under ideal noise conditions. By ``ideal noise" we mean the noise satisfies the assumptions above for the method of~\cite{ryan_inprep2013} to hold. The details of this calculation, along with a brief introduction to measurements and amplifiers, is contained in~\cite{Supp}. We obtain
\begin{align}
\mathcal{F}_\mathrm{ach} &= 0.9999 \pm 0.0001,
\end{align}
and so there is a large discrepancy between $\mathcal{F}_{\mathrm{ach}}$ and $\mathcal{F}_a$ that is due to a wide variety of factors such as state-preparation errors and non-Gaussian/non-linear effects. This discrepancy provides the motivation for investigating better methods for classifying trajectories.

The idea behind machine learning (ML) classification is to obtain a classifying (discriminating) surface in $\mathbb{R}^M$ under constraints such as the form of the noise. For Gaussian noise the optimal discriminator is a quadratic surface~\cite{Cover65} (quadratic discriminant analysis-QDA) given by
\begin{align} \label{eq:QDAeq}
f_{\text{QDA}}(x)&= -\frac{1}{2}x^T\left[\Sigma_0^{-1}-\Sigma_1^{-1}\right]x + x^T \left[\Sigma_0^{-1}\mu_0-\Sigma_1^{-1}\mu_1\right],
\end{align}
and the threshold is
\begin{align} \label{eq:QDAthresh}
v_{\text{QDA}} = \frac{1}{2}\left(\mu_0^T\Sigma_0^{-1}\mu_0 - \mu_1^T\Sigma_1^{-1}\mu_1\right) + \frac{1}{2} \log\left(\frac{|\Sigma_0|}{|\Sigma_1|}\right),
\end{align}
where $| \cdot |$ represents the determinant. If $\Sigma_0 = \Sigma_1$ the quadratic term in Eq.~(\ref{eq:QDAeq}) disappears and the surface reduces to a hyperplane~\cite{Fisher36} (linear discriminant analysis-LDA),
\begin{align}
f_{\text{LDA}}(x)&= x^T\left[\Sigma^{-1}(\mu_0-\mu_1)\right].
\end{align}
Comparing with Eq.~(\ref{eq:LDAdiag}) we see the current method of~\cite{ryan_inprep2013} is equivalent to LDA with the added assumption of diagonal covariance matrices. 


QDA can achieve a more accurate value of $\mathcal{F}_a$ as it allows for  $\Sigma_0 \neq \Sigma_1$ and thus a quadratic (non-linear) discriminating surface. We computed $\mathcal{F}_a$ using the ``fitcdiscr" function in Matlab for four different methods: LDAd, LDA, QDAd, and QDA (``d" represents diagonal covariance matrix and LDAd is the method of~\cite{ryan_inprep2013}). The results are in the second column of Table~\ref{table:Assfid_discriminant_methods}. Not surprisingly, we find QDAd improves upon LDAd and allowing non-diagonal covariance matrices produces higher $\mathcal{F}_a$. The values in Table~\ref{table:Assfid_discriminant_methods} are the sample means from 100 repetitions. The sample variances $\sigma^2$ are typically on the order of $1\times 10^{-8}$ indicating stable and reproducible results.

A value of $\mathcal{F}_a$ for QDA was not attainable due to singular covariance matrices, which is a result of overfitting the data (having more variables than required from the correlation time in the trajectories). To remedy this problem, we first perform dimensionality reduction using principal component analysis (PCA)~\cite{Pearson01} and find $99.9\%$ of the variance in the data can be accounted for in a subspace of dimension $\sim 20$. Loosely speaking, this implies correlation times of $\sim 85-180$ ns. The results with a PCA pre-processing step (using ``princomp" in Matlab) are in the third column of Table~\ref{table:Assfid_discriminant_methods}. A value of QDA can now be computed and as expected it provides the highest $\mathcal{F}_a$ out of all cases considered.

\begin{table}[ht]
\caption{Assignment fidelities for various discriminant analysis methods. See text for details.} 
\centering 
\begin{tabular}{c c c c c} 
\hline\hline 
Method & All time-points & PCA \\ [0.5ex] 

\hline 
LDAd & 0.9586 & 0.9557 \\
LDA & 0.9701 & 0.9586\\
QDAd & 0.9627 & 0.9648\\
QDA & -- & 0.9712\\ [1ex]
\hline 
\end{tabular}
\label{table:Assfid_discriminant_methods}
\end{table}

These classification methods have assumed Gaussian noise. More robust methods are needed as we expect non-Gaussian behavior. We approach this in two ways. The first is via the support vector machine (SVM)~\cite{Boser92,Cortes_Vapkin_95}, which makes no assumption on the form of the noise and can be extended to extremely general non-linear discriminating surfaces. The second is to utilize ``clustering" methods in ML to naturally goup the data into clusters from which we perform \emph{multi-class} classification. We first describe the SVM method.

The linear SVM is a quadratic program based on maximizing the minimum margins of the data, where the margin of a data point is its distance to a separating hyperplane~\cite{Supp}. The non-linear SVM is derived from the dual form of the linear SVM by defining a kernel that maps the data to a higher-dimensional space. The key is that the linear SVM in the higher-dimensional space allows for non-linear discrimination in $\mathbb{R}^M$. Due to its generality and simplicity, we chose a radial basis (Gaussian) function kernel. We implemented the SVM using the ``fitcsvm" function in Matlab. The classification was repeated 100 times and the mean values with the optimal soft-margin parameter are contained in Table~\ref{table:Assfid_SVM_methods} (see~\cite{Supp} for details). The sample variances $\sigma^2$ in $\mathcal{F}_a$ are approximately $1.9 \times 10^{-8}$ indicating stable results. The non-linear SVM produces the highest assignment fidelity out of all methods considered thus far, indicating non-linear effects are present.

\begin{table}[ht]
\caption{Assignment fidelities for SVM methods. See text for details.} 
\centering 
\begin{tabular}{c c c c c} 
\hline\hline 
Method & All time-points & PCA \\ [0.5ex] 

\hline 
Linear SVM & 0.9753 & 0.9571 \\
Non-linear SVM & 0.9821 &  0.9739 \\ [1ex]
\hline 
\end{tabular}
\label{table:Assfid_SVM_methods}
\end{table}

Our second method for implementing a non-linear classifier combines classification and \emph{clustering} algorithms. Clustering naturally groups the data into subsets and is ``unsupervised" since it requires no training data. We utilize k-means clustering~\cite{Lloyd82} since it has features that suit our purposes well, however we anticipate similar results can be obtained with other standard clustering algorithms such as heirarchical methods. For an explicit formulation of the k-means clustering problem see~\cite{Supp}.

\begin{figure}[h!]
\centering
\includegraphics[trim = 0.5in 2.0in 0.1in 2.5in,width=0.43\textwidth]{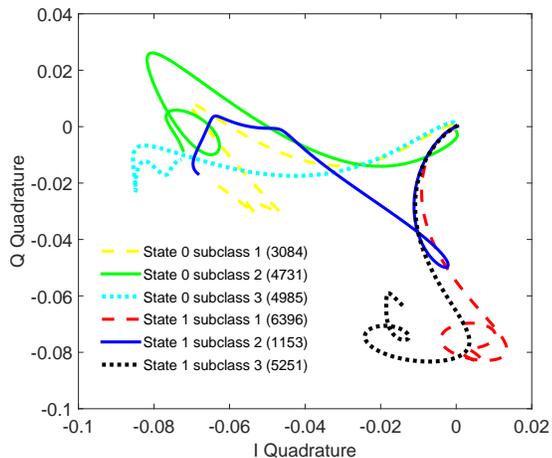}
\caption{\label{fig:Subclass_M4} Subclasses found from k-means algorithm (color online). $C_0$ and $C_1$ have three subclasses, the trajectory representing each subclass is the mean over all subclass trajectories. The subclasses of $C_0$ (yellow-dashed, green-solid, light blue-dotted) have paths that initially move up and left. The subclasses of $C_1$ (red-dashed, blue-solid, black-dotted) have paths that initially move down and right. The $\mathrm{T}_1$ subclass (blue-solid) of $C_1$ initially moves down and right but abruptly changes its path to move up and left. The legend numbers are subclass sizes.}
\end{figure}

We used the Matlab ``kmeans" function to find $k=3$ clusters in each of $C_0$ and $C_1$. We chose $k=3$ to take into account both variance and systematic effects. The mean trajectories and size of the six subclasses are given in Fig.~\ref{fig:Subclass_M4}. We see $C_0$ is split relatively evenly into the subclasses $S_{0,1}$, $S_{0,2}$, $S_{0,3}$ that capture variance in the trajectories. We do not see a subclass of $C_0$ corresponding to heating of the ground state, however we implemented k-means for larger $k$ and found a heating subclass of size $\sim 230$ for $k=7$ (see Fig.~\ref{fig:Heating_subclass}).

 $C_1$ has strikingly different properties as subclass $S_{1,2}$ is comprised of $\mathrm{T}_1$ processes. $S_{1,1}$ and $S_{1,3}$ are similar in size and capture variance in the trajectories. The key point is we have found \emph{explicit shot indices} for $\mathrm{T}_1$ events. We verified that $S_{1,2}$ is comprised of $\mathrm{T}_1$ trajectories by performing k-means with $k=4$. We found that the $S_{1,1}$ and $S_{1,3}$ subclasses remain virtually the same while the $\mathrm{T}_1$ subclass $S_{1,2}$ is now split into two according to variance in these trajectories (see Fig.~\ref{fig:Four_subclasses}). From Fig.~\ref{fig:Subclass_M4}, $\sim 9\%$ of the $|1\rangle$ preparations result in a $\mathrm{T}_1$ event, which is consistent with the percentage calculated from system parameters~\cite{Supp}, $1-e^{-2.6/29} \sim 8.6\%$.

To perform classification, we lift the $\mathrm{T}_1$ subclass $S_{1,2}$ to a class $C_2$ of its own, redefine $C_1 = S_{1,1} \cup S_{1,3}$, keep $C_0$ as before, and perform \emph{multi-class} classification on $C_0$, $C_1$, and $C_2$. We implemented four multi-class algorithms in Matlab; multi-class LDA, multi-class SVM, ``TotalBoost", and ``RUSBoost". The latter two are examples of boosting algorithms which assemble an ensemble of weak learners (classifiers) in a network to create a final strong learner by iteratively re-weighting data points according to previous results~\cite{Bishop07}. The RUSBoost method~\cite{Seiffert10}  is particularly useful since it is tailored to the case of one class (here $C_2$) being significantly smaller than the rest. 

The results are in Table~\ref{table:Assfid_multiclass_methods}. We again see an increase in assignment fidelities over the discriminant analysis methods of Table~\ref{table:Assfid_discriminant_methods}. Not surprisingly, RUSBoost provides the most significant increase. We repeated the k-means algorithm 50 times with random initializations and found it to be relatively stable (sample variance $\sigma^2$ of $\mathcal{F}_a \sim 3 \times 10^{-6}$). We repeated this using fixed initialization of the means and obtained a variance of 0.

Out of all methods considered, non-linear SVM's produce the greatest increase in $\mathcal{F}_a$ (0.9586 to 0.9821). We also note all methods are relatively stable with reproducible assignment fidelities (each method was repeated $\sim 100$ times; the sample means of $\mathcal{F}_a$ are the table values and the sample variances are $\sim 1\times 10^{-8}$).

\begin{table}[ht]
\caption{Assignment fidelities from multi-class classification. See text for details.} 
\centering 
\begin{tabular}{c c c} 
\hline\hline 
Method & All time-points & PCA \\ [0.5ex] 

\hline 
Multi-LDA & 0.9768 & 0.9689\\
Multi-SVM & 0.9784 & 0.9717 \\
TotalBoost & 0.9527 & 0.9413  \\
RUSBoost & 0.9788 & 0.9723 \\[1ex]
\hline 
\end{tabular}
\label{table:Assfid_multiclass_methods}
\end{table}

While we have improved $\mathcal{F}_a$ to 0.9821, we are still far from $\mathcal{F}_\mathrm{ach} = 0.9999$. It is possible much of the remaining discrepancy comes from $\mathrm{T}_1$ events. To investigate this we propose the simple error diagnosis test of replacing each $\mathrm{T}_1$ event found from the k-means algorithm with a random element from $S_{1,2} \cup S_{1,3}$. This provides a measure of $\mathcal{F}_a$ when $\mathrm{T}_1$ is negligible. The means of 100 samples for each method are contained in Table~\ref{table:Assfid_replace} (variances are $\sim 1 \times 10^{-8}$). Non-linear SVM produces the highest value of $\mathcal{F}_a$ however \emph{for all methods} $\mathcal{F}_a > 0.99$, which is more consistent with $\mathcal{F}_\mathrm{ach} = 0.9999$. This confirms $\mathrm{T}_1$ events are the significant reason for lower $\mathcal{F}_a$ values than expected.

\begin{table}[ht]
\caption{Assignment fidelities with replacement of $\mathrm{T}_1$ events. See text for details..} 
\centering 
\begin{tabular}{c c c} 
\hline\hline 
Method & All time-points & PCA \\ [0.5ex] 

\hline 
LDAd & 0.9920 & 0.9909\\ 
LDA & 0.9921 & 0.9928 \\
QDAd & 0.9918 & 0.9908 \\
QDA & -- & 0.9927 \\
Linear SVM & 0.9936 & 0.9943 \\
Non-linear SVM & 0.9945 & 0.9949 \\[1ex]
\hline 
\end{tabular}
\label{table:Assfid_replace}
\end{table}

\begin{figure}[h!]
\centering
\includegraphics[trim = 0.5in 2.3in 0.1in 2.5in,width=0.43\textwidth]{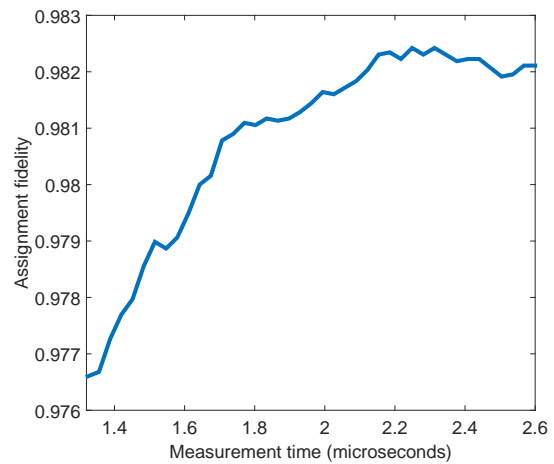}
\caption{\label{fig:Assfid_vary_time} Varying measurement time.}
\end{figure}

One attempt to reduce the significance of $\mathrm{T}_1$ is to reduce $T$, however this implies the trajectories will spend less time near their steady states and assignment errors due to variance will increase.To observe this, we truncated the trajectories to different $T$ and calculated $\mathcal{F}_a$ using the non-linear SVM. From Fig.~\ref{fig:Assfid_vary_time} we see $T=2.6\mu$s appears close to optimal. Moreover, a much shorter measurement time of $\sim 1.2 \mu$s (not shown in Fig.~\ref{fig:Assfid_vary_time}) is needed to achieve $\mathcal{F}_a$ from LDA. This is a strong message that better classifiers can allow for shorter measurement times. Longer measurement times than the current $2.6 \mu$s decrease $\mathcal{F}_a$ due to an increase in $\mathrm{T}_1$ events.

To conclude, we have utilized ML to understand and improve the readout in a superconducting system. We find more sophisticated classification algorithms can potentially allow for shorter measurement times and increase assignment fidelities. Non-linear SVM's provided the largest increase in assignment fidelity from 0.9586 to 0.9821 ($\sim 2.4\%$). Clustering helped diagnose the prevalence of systematic effects by finding clusters in the data corresponding to single-shot identification of heating and $\mathrm{T}_1$ effects. We verified $\mathrm{T}_1$ events are a significant source of error as the assignment fidelity increases from 0.9821 to 0.9945 when the $\mathrm{T}_1$ cluster is replaced with typical trajectories. This is more consistent with our achievable fidelity and the remaining discrepancy can be due to effects such as heating and state-preparation errors. Moving forward, we expect these methods will help provide insight for improving readout, especially when non-linear and non-Gaussian effects are present.

\begin{acknowledgments}
We acknowledge support from
ARO under contract W911NF-14-1-0124 and  IARPA under contract W911NF-10-1-0324. We acknowledge helpful discussions with Oliver Dial, Stefan Filipp, Blake Johnson, Jim Rozen, Colm Ryan, Marcus Silva, and Matthias Steffen.
\end{acknowledgments}


\newpage

\section{Supplemental Material}

\subsection{Measurement and linear amplification in circuit QED}\label{sec:amp}

In the dispersive regime of circuit-QED the resonator frequency depends on the qubit state so that driving the cavity and observing the output of the cavity corresponds to a quantum non-demolition measurement. For outcome ``0" (1) the mean voltage leaving the cavity represents a time-dependent coherent state, denoted $\alpha_{0 (1)}(t)$, that is typically small in magnitude. The phase-space evolution of $\alpha_{0 (1)}(t)$ is determined by the deterministic differential equation
\begin{align}
\dot{\alpha}_{0(1)} &= -i\mathcal{E}(t) - i\left(\omega_r + \chi_{0(1)}\right)\alpha_{0(1)} - \left(\kappa \alpha_{0(1)}\right)/2,
\end{align}
where $\mathcal{E}(t) = \mathcal{E}_x(t)\cos(\omega_m t) + \mathcal{E}_y(t)\sin(\omega_m t)$ is the measurement drive and $\kappa$ is the cavity decay rate.

For our experiment, the qubit transition frequency is $\omega/2\pi= 5.415$ GHz and the readout resonator frequency is $\omega_{R}/2\pi= 6.693$ GHz. The qubit anharmonicity is  $\sim 330$ MHz with $\mathrm{T}_1$ (energy relaxation) time of $29 \mu$s and coherence time $T_2^{\rm{echo}} = 22 \mu$s. The dispersive shift and line width of the readout resonator are measured to be $2\chi/2\pi  = -2.8$ MHz and $\kappa/2\pi = 1210$ kHz, respectively so that $\chi \sim \kappa$.

The output mode $b(t)$ from the resonator can be related to the field inside the resonator by
 \begin{align}\label{eq:field}
  \left \langle b(t)\right \rangle = \sqrt{\kappa } \frac{\beta(t)}{2}\langle z \rangle  + \sqrt{\kappa} \frac{\nu(t)}{2},
 \end{align} where $\beta(t) =\alpha_0(t)  - \alpha_1(t)$ is the separation between the pointer states, $\nu(t)=\alpha_0(t)  + \alpha_1(t)$ is the mean value of the coherent states and $\langle z \rangle = p_0-p_1$ represents the qubit information with $p_0$ and $p_1$ being the probability of the qubit being in state 0 and 1 respectively~\cite{gambetta_trajectories_2008}. Typically this field is amplified by a 
linear phase-preserving amplifier to the output mode $c(t)$,
\begin{align}\label{eq:outputmode}
c(t)&=\sqrt{G}b(t)+h(t),
\end{align} where $G$ is the power gain and $h(t)$ is the extra noise added by the amplifier. Here we have made an assumption that the bandwidth of the amplifier is constant and larger than the bandwidth of the signal being measured.  Since the commutators must satisfy
\begin{align}\label{eq:commutation}
\left[c(t),c^{\dagger}(t^\prime)\right] &= \left[b(t),b^{\dagger}(t^\prime)\right] = \openone \delta(t-t^\prime),
\end{align}
the noise must satisfy
\begin{align}\label{eq:noisecommutation}
\left[h(t),h^\dagger(t')\right] = \openone(1-G)\delta(t-t'),
\end{align}  
which, from the generalized uncertainty principle of an arbitrary operator $R$,
\begin{align}
|\Delta R|^2 \ge \frac{1}{2}|\langle[R,R^\dagger]\rangle|
\end{align}
implies $
|\Delta h|^2\ge \frac{1}{2}|1-G| \delta(t-t')$ and $|\Delta b|^2\ge \frac{1}{2}\delta(t-t')$. Using Eq. \ref{eq:outputmode} the noise in the output mode $c(t)$ is 
\begin{align}
|\Delta c|^2\ge \frac{G(1+2A)\delta(t-t')}{2},
\end{align} where 
\begin{align}\label{ea:Anoise}
A\ge\frac{1}{2}\left(1-\frac{1}{G}\right)
\end{align}is the added noise normalized by the gain~\cite{Caves82}. In our view this is the best number to quantify an amplifier and for the quantum limit it takes the value 1/2 for a phase preserving amplifier. Other useful quantities are the instantaneous input and output signal-to-noise ratios of the amplifier defined by
\begin{align}
\mathrm{SNR}_\mathrm{out} & =  \frac{|\langle c\rangle |^2 }{|\Delta c|^2} = {\eta_l|\langle b\rangle|^2 \delta t} \\
\mathrm{SNR}_\mathrm{in} &= \frac{|\langle b\rangle|^2}{|\Delta b|^2}=2|\langle b\rangle |^2 \delta t,
\end{align} where $\eta_l = 2/(1+2A)\le 1 $ is the efficiency of the amplifier and represents how well the input field is mapped to the output field.  Another useful quantity is the noise-figure $f$ of the amplifier, which is the ratio $\mathrm{SNR}_\mathrm{in}/\mathrm{SNR}_\mathrm{out}$. We see that $f=1+2A$ which is 2 for a quantum limited amplifier.

In circuit QED the information about the qubit state is contained in a single quadrature. From Eq. \ref{eq:field} this is the quadrature set by $\theta = \mathrm{arg}(\beta(t))$ and as a result when we subtract the mean value we obtain
\begin{align}
\mathrm{SNR}_\mathrm{out} & =\frac{\kappa \eta_l|\beta(t)|^2 \delta t}{4}.
\end{align} This overestimates the noise as the information is only in one quadrature. Defining the measurement quadrature $I(t) = \mathrm{Re}[c(t) e^{-i\theta}]$
 it is simple to show that 
\begin{align}
|\Delta I|^2 \ge \frac{1}{2}|\Delta c|^2  = \frac{G}{4}(1+2A)\delta(t-t'),
\end{align} giving an instantaneous SNR$_I$
\begin{align}
\mathrm{SNR}_I & = \frac{\langle I(t)\rangle ^2}{|\Delta I|^2 } = \eta \kappa |\beta(t)| ^2\delta t,
\end{align} where $\eta = 1/(1+2A) \leq \frac{1}{2}$ is the efficiency of measuring information in a single quadrature for a linear phase preserving amplifier.  
Note this is a factor of two less than the efficiency of the amplifier. 

A general linear amplifier can be described by the output mode $c(t)$,
\begin{align}\label{eq:outputmode2}
c(t)&=M b(t)+L b^\dagger(t)+h(t),
\end{align} and from preservation of the commutation relations 
\begin{align}\label{eq:noisecommutation2}
\left[h(t),h^\dagger(t')\right] = \openone(1-M^2+|L|^2)\delta(t-t'),
\end{align} we have
$|\Delta h|^2\ge \frac{1}{2}|(1-M^2+|L|^2)| \delta(t-t')$. Setting $|L|=\sqrt{M^2-1}$ results in  $|\Delta h|^2\ge 0$. To amplify a single quadrature $I = \mathrm{Re}[c(t) e^{-i\theta}]$ (phase sensitive amplifier) we set $L= \sqrt{M^2-1}e^{2i\theta}$ giving 
\begin{align}
I(t)&= \sqrt{G} \mathrm{Re}[ b(t)e^{-i\theta} ] +\mathrm{Re}[ h(t)e^{-i\theta}],\\
Q(t)&=\frac{1}{\sqrt{G}}  \mathrm{Im}[ b(t)e^{-i\theta} ] +\mathrm{Im}[ h(t)e^{-i\theta}],
\end{align} where $G=(M+\sqrt{M^2-1})^2$ is the power gain. From the generalized uncertainty principle 
 \begin{align}
|\Delta I|^2 \ge\frac{ G_s}{4}(1+2 A_s)\delta(t-t'),
 \end{align} where $A_s\ge0$ is the gain normalized added noise. The instantaneous SNR$_I$ for the quadrature $I$ is
 \begin{align}
 \mathrm{SNR}_I & = \frac{\langle I(t)\rangle ^2}{|\Delta I|^2 } = \eta \kappa |\beta(t)| ^2\delta t.
 \end{align} where $\eta = 1/(1+2 A_s) \leq 1$. That is, by using a phase sensitive amplifier tuned to the correct phase the effective SNR can be a factor of two better then a phase preserving amplifier.

\subsection{Filtering protocol for ideal noise and the achievable fidelity}

In a typical measurement protocol the measurement outcome is the integration of the signal $c(t)$ from the amplifier with a weighting kernel (filter),  
\begin{align}
S =  \int_0^{t_m}\mathrm{Re}[w(t)c(t) ]dt = \sum_j |w_j|\mathrm{Re}[ e^{-i\phi_j }c_j ] \delta t,
\end{align} 
where $t_m$ is the measurement time and  $w_j=|w_j|e^{-i\phi_j }$ is the kernel. Under the assumption that the noise is symmetric a useful measure  to quantify the measurement is the separation 
\begin{align}\label{eq:sep}
R = \frac{(\langle S_0\rangle -\langle S_1\rangle)^2}{\mathrm{var}(S)},
\end{align} where 0 and 1 label the two states of the qubit.  In Ref.~\cite{ryan_inprep2013} it was shown that the optimal kernel under the additional assumptions of Gaussian and diagonal covariance matrices is found by maximizing $R$ and is given by
\begin{align}
w_j = \frac{\beta_j^*}{\mathrm{var}(I_j)},
\end{align} where $\beta_j$ is the separation of the pointer states and $I_j = \mathrm{Re}[e^{-i\theta_j }c_j]$.

\begin{figure}[h!]
 \centering
 \includegraphics[trim = 0.5in 2.0in 0.1in 2.0in,width=0.43\textwidth]{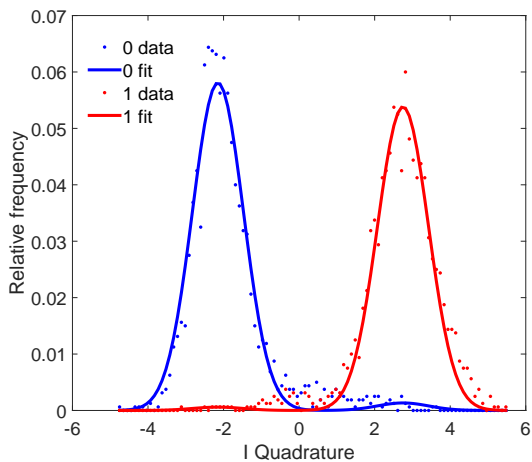}
 \caption{\label{fig:Gaussian_fits} Projected data and double Gaussian fits of the $|0\rangle$ (blue-left) and $|1\rangle$ (red-right) preparation classes (color online).}
 \end{figure}

The achievable fidelity $\mathcal{F}_\mathrm{ach}$ is the value of the assignment fidelity $\mathcal{F}_a$ assuming that the noise in our system is ideal (symmetric and diagonal Gaussian covariance matrices). In reality, the noise does not satisfy these properties so we must fit it to an ideal noise model. To do this, we implemented the above filtering method on our data and plotted the resulting distributions for the different classes ($|0\rangle$ and $|1\rangle$) of $S$ in Fig.~\ref{fig:Gaussian_fits}. If the noise was ideal we should obtain two Gaussian histograms with identical variance. In reality, we see there are significant non-Gaussian statistics and so we fit the histograms to double Gaussian distributions of equal variance to obtain the means and standard deviation we would expect in the ideal noise case. We obtain $\langle S_0\rangle=-0.2149 \pm 0.0149$ and $\langle S_1\rangle=2.750\pm 0.0161$, and a standard deviation of $\sqrt{\mathrm{var}(S)}=0.6758 \pm 0.0109$, which gives a value for $R$ (defined in Eq.~\ref{eq:sep}) of $55.56 \pm 1.02$.

We can now compute $\mathcal{F}_\mathrm{ach}$ by combining the ideal noise assumption with the definition of the assignment fidelity,
\begin{equation}\label{eq:assfid}
\mathcal{F}_a = 1-\left(\mathbb{P}\left[0|1\right] + \mathbb{P}\left[1|0\right]\right)/2.
\end{equation}
Here $\mathbb{P}[0|1]$ ($\mathbb{P}[1|0]$) is the probability of obtaining outcome ``1" (``0") given the system was prepared in $|0\rangle$ ($|1\rangle$). Since the noise is ideal $\mathcal{F}_\mathrm{ach}$ takes the form
\begin{equation}\label{eq:assfid2}
\mathcal{F}_a = 1/2+\mathrm{erf}(\sqrt{R/8})/2,
\end{equation} where $R$ is defined in Eq. \ref{eq:sep}.  From this expression we obtain 
 \begin{align}
 \mathcal{F}_\mathrm{ach} &= 0.9999 \pm 0.0001
 \end{align} for our system. 
We note that the separation $R$ in the time independent limit can be related to the signal-to-noise defined above by SNR$_I=R/4$. This is straightforward to show by direct substitution of the quantities in Eq. \ref{eq:sep}. $R$ is used here as it is a standard measure of separation in ML~\cite{Fisher36}.

\subsection{SVM's and k-means clustering}

The quadratic program for the SVM~\cite{Cortes_Vapkin_95} is given by the equation
\begin{equation}\label{eq:SVMprimal}
\begin{aligned}
& \underset{w,b}{\text{minimize}}
& & \|w\|^2/2 \\
& \text{subject to}
& & y^{(i)}(w^Tx^{(i)} + b)   \geq 1, \: \: i \in \{1,...,M\},
\end{aligned}
\end{equation}
where $y$ is the expected outcome (taken to be -1 or 1). This has a quadratic objective function with linear inequality constraints. The soft-margin formalism adds slack variables representing the degree of misclassification to the constraints in Eq.~(\ref{eq:SVMprimal}) and modifies the objective function to include a mislabelling cost term.
The modified quadratic program that includes a soft-margin is given by the equation
\begin{equation}\label{eq:SVMprimalsoft}
\begin{aligned}
& \underset{w,b,\xi}{\text{minimize}}
& & \|w\|^2/2 + C\sum_{j=1}^M \xi_j \\
& \text{subject to}
& & y^{(i)}(w^Tx^{(i)} + b)   \geq 1-\xi_i, \: \: i \in \{1,...,M\}.
\end{aligned}
\end{equation}
In the dual form of this problem, the $\xi$ variables vanish and $C$ is a ``box constraint" that bounds the Lagrange multipliers.  We implemented 10-fold cross-validation on the training set and found a misclassification error of 0.0163 which agrees well with $\mathcal{F}_a = 0.9825$. This implies large $C$ value is likely not needed and varying $C$ between 0 and 100 gave an optimal value of $\mathcal{F}_a$ close to 1.

k-means clustering~\cite{Lloyd82} is formulated as the following optimization problem
\begin{equation}\label{eq:kmeans}
\begin{aligned}
& \underset{\bf{S}}{\text{minimize}}
& & \sum_{j=1}^k \sum_{x \in S_j} \| x - \mu_j\|^2.
\end{aligned}
\end{equation}
Here there are $k$ sets $S = \{S_1,...,S_k\}$ with means $\mu_j$ and so the goal is to partition the data into $k$ sets that minimize the within-set distance to the mean $\mu_j$. The k-means algorithm only has guaranteed convergence to a local minimum and finding the global optimum is an NP-hard problem~\cite{Aloise09}. Therefore initialization of the subclass means in Eq.~\ref{eq:kmeans} can be important to find meaningful solutions. Typically initialization is a random process that can lead to small variation in the solutions. This can be circumvented by explicitly defining the initial means to be the average of the output subclass means over many realizations. This helps ensure the algorithm is reproducible and more stable. 

\begin{figure}[h!]
 \centering
 \includegraphics[trim = 0.5in 2.0in 0.1in 2.0in,width=0.43\textwidth]{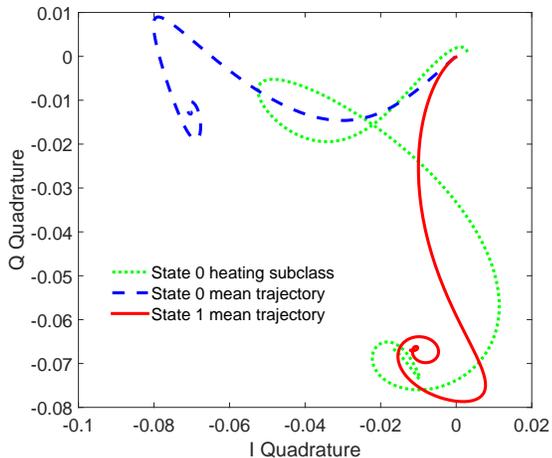}
 \caption{\label{fig:Heating_subclass} Heating subclass of $C_0$ (green-dotted) found from a k-means algorithm with $k=7$ superimposed on means of $|0\rangle$ (blue-dashed) and $|1\rangle$ (red-solid) classes (color online).}
 \end{figure}

\begin{figure}[h!]
 \centering
 \includegraphics[trim = 0.5in 2.0in 0.1in 2.0in,width=0.43\textwidth]{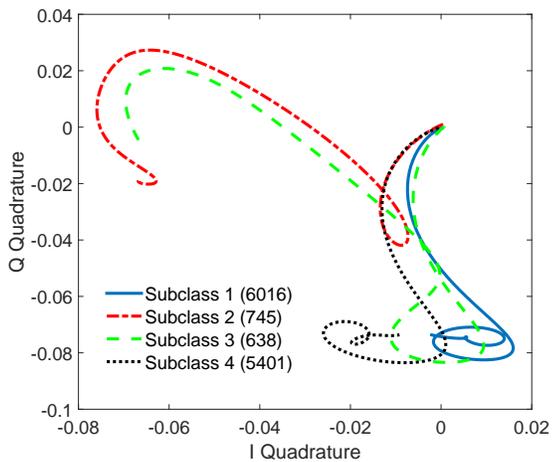}
 \caption{\label{fig:Four_subclasses} Four subclasses of $C_1$ found from a k-means algorithm. There are two $\mathrm{T}_1$ subclasses (green-dashed and red-dash-dotted) that split the subclass found from $k=3$. The other subclasses (black-dotted and blue-solid) are comprised of typical trajectories of a $|1\rangle$ state (color online).}
 \end{figure}

The plot of the heating subclass of $C_0$ found from the k-means algorithm with $k=7$ is given in Fig.~\ref{fig:Heating_subclass} and the plot of the four subclasses of $C_1$ found from the k-means algorithm with $k=4$ is given in Fig.~\ref{fig:Four_subclasses}.

%

\end{document}